%% file: main.tex
\documentclass[a4paper,parskip=half]{scrartcl}
\pdfoutput=1
\usepackage{graphicx}
\usepackage{mystyle}
\usepackage[colorlinks,
	    urlcolor    = Venetian,
	    citecolor   = Lime,
	    linkcolor   = Dodger,
	    linktocpage = true,
	    ]{hyperref}
\hypersetup{
            pdftitle={Reweighting a parton shower using a neural network: the final-state case},
            pdfauthor={Enrico Bothmann; Luigi Del Debbio},
            pdfkeywords={qcd, parton shower, sudakov veto algorithm, reweighting, machine learning},
            pdfborder={0 0 0},
            breaklinks=true}
\title{Reweighting a parton shower using a neural network: the final-state case}
\author{Enrico Bothmann\thanks{\href{mailto:enrico.bothmann@ed.ac.uk}{enrico.bothmann@ed.ac.uk}}\ }
\author{Luigi Del Debbio}
\affil{University of Edinburgh}
\date{}

\begin{document}

\setcounter{secnumdepth}{2}
\setcounter{tocdepth}{2}
\maketitle
\input{text/abstract}
\tableofcontents

\input{text/intro}

\section{Reweighting the Sudakov Veto Algorithm}
\label{sec:rew}
\input{text/rew_sva}
\input{text/ps_reweighting}

\section{Neural-network approach to a-posteriori parton-shower reweighting}
\label{sec:nn}
\input{text/shower_model}
\input{text/nn_io}
\input{text/nn_arch_and_training}

\section{Validation}\label{sec:validation}
\input{text/validation_procedure}
\input{text/results}

\input{text/example}

\input{text/conclusions}

\ifdraft{\input{text/sizes}}{}

\bibliography{refs}
\bibliographystyle{utphys}

\end{document}

%% file: text/abstract.tex
\begin{abstract}
\noindent
The use of QCD calculations that include the resummation of soft-collinear
logarithms via parton-shower algorithms is currently not possible in PDF fits
due to the high computational cost of evaluating observables for each
variation of the PDFs. Unfortunately the interpolation methods that are otherwise applied
to overcome this issue are not readily generalised to all-order parton-shower
contributions. Instead, we propose an approximation based on training a neural
network to predict the effect of varying the input parameters of a parton shower on
the cross section in a given observable bin, interpolating between the
variations of a training data set. This first publication focuses on providing a
proof-of-principle for the method, by varying the shower dependence on $\alphas$
for both a simplified shower model and a complete shower implementation for
three different observables, the leading emission scale, the number of emissions
and the Thrust event shape. The extension to the PDF dependence of the
initial-state shower evolution that is needed for the application to PDF fits is
left to a forthcoming publication.
\end{abstract}

%% file: text/intro.tex
\section{Introduction}

With the Large Hadron Collider (LHC) successfully undergoing its second run and
the possible upgrade to the so-called High-Luminosity LHC, the produced
collision datasets reach new levels of precision. This requires an ongoing
effort to provide more precise theory predictions. A sizeable part of the
overall theory uncertainty is often given by the degree to which we know the
parton content of the incoming protons, parametrised by the parton density
functions
(PDF)~\cite{Ball:2017nwa,Dulat:2015mca,Harland-Lang:2014zoa,Alekhin:2013nda},
see also Ref.~\cite{Gao:2017yyd} and references therein for a recent review. In
order to increase the precision of PDF determinations, it is clearly desirable
to be able to include as many observables as possible in the fits. However the
$\chi^2$ minimization in these fits needs multiple re-evaluations of the
underlying observables in order to converge. As a result there are strict
constraints on the CPU time each re-evaluation costs: observables can be
included in fits only if there is an efficient way to compute them as PDFs are
varied. For instance, in \textsc{nnpdf} fits, all observables are written as
convolutions
of FK tables and PDFs at the initial scale as discussed in detail in
Ref.~\cite{Ball:2014uwa}. Similarly, some results provided by Monte-Carlo event
generators, e.g.\ \mbox{\textsc{NLOJet\texttt{++}}}~\cite{Nagy:2003tz},
\textsc{MadGraph5\_aMC@NLO}~\cite{Alwall:2014hca},
\textsc{mcfm}~\cite{Campbell:2015qma} or
\textsc{sherpa}~\cite{Gleisberg:2008ta}, can be projected onto an interpolation
grid, which allows fast a-posteriori variations of the input parameters, because
the sum over simulated events is replaced by a sum over a much smaller number of
observable-dependent
weights~\cite{Carli:2010rw,Kluge:2006xs,Britzger:2012bs,DelDebbio:2013kxa,
Bothmann:2015dba,Bertone:2014zva}. Unfortunately, as explained later in this
work, this approach cannot be easily extended to capture the input-parameter
dependences of the all-order predictions obtained by the parton-shower algorithms
in Monte-Carlo generators.

Instead we present here an approximate approach to parametrise the parton-shower
dependences in a way that allows for a fast, a-posteriori reweighting of the
observable. Because we are dealing with binned observables, the output of the
parton shower is the number of events in a given bin. In this context
'reweighting the observable' means finding the relative weight of each bin as
the input parameters change, i.e. recomputing the value of the observable in
each bin. Once the relative weight is known, the number of events in a bin can
be easily recomputed by multiplying its original value by the new weight. A
quick summary of the method is as follows. First, the calculation is repeated
for a given set of input parameters. The variation of the result (in a given
observable bin) across this set is then used to train a neural network (NN),
effectively fitting the unknown functional form that encodes the dependences of
the parton shower on the input parameters. This NN can then be used to obtain
efficiently an interpolation of the observable for arbitrary values of the input
parameters, so that it is suitable to use this methodology in studies that
require fast a-posteriori variations. 

NN techniques have been successfully applied or used exploratively in a number
of topics in collider phenomenology, often with much more complex NN
architectures than what we employ in this work. The topics include jet
tagging/particle
identification~\cite{deOliveira:2015xxd,Lin:2018cin,Macaluso:2018tck,
Butter:2017cot,Luo:2017ncs,Cheng:2017rdo,Louppe:2017ipp,Komiske:2016rsd,
Shimmin:2017mfk},
event classification~\cite{Nguyen:2018ugw,Metodiev:2017vrx,Abdughani:2018wrw},
phase-space
integration~\cite{Bendavid:2017zhk}, pile-up mitigation~\cite{Komiske:2017ubm},
simulating electromagnetic showers in a calorimeter~\cite{Paganini:2017dwg},
parameter space scans for New Physics searches~\cite{Ren:2017ymm}, and
of course PDF fitting~\cite{Ball:2017nwa}. Moreover, a deep NN has been proposed to
mimic a parton shower algorithm~\cite{Monk:2018zsb}. However, this latter ansatz
can not be applied to our goal of using all-order results in PDF fits, as it is
applied on an event-by-event basis just as an ordinary parton-shower algorithm,
whereas PDF fits require projections of the cross section on observables in
order to achieve the fast evaluation times needed for the fit. The interplay of
parton showers and NN has also been studied in~\cite{Barnard:2016qma}, in which
the authors investigate the role of parton-shower uncertainties and
approximations in NN-based jet substructure analyses.
In~\cite{Carrazza:2018mix} an NN is used to determine the effective correction
to a Sudakov form factor in a \textsc{minlo} calculation of single-top plus jet
production.

In this exploratory study, we restrict ourselves for simplicity to final-state
parton showers. While these are not dependent on PDFs, the problems that need to
be addressed are the same that would appear in a PDF fit. This simplified
setting allows us to test our ideas without getting bogged down in
technicalities. The same approach can be readily generalised to include
PDF-dependent initial-state emissions. In the latter case, having to deal with a
much larger space of parameters entails a number of algorithmic issues that will
be addressed in future studies.


In Sec.~\ref{sec:rew}, we summarise the main steps in the algorithm for
reweighting a parton shower on-the-fly, highlighting the fact that a
straightforward generalisation of the interpolation approach used e.g. in
Ref.~\cite{DelDebbio:2013kxa} does not seem feasible. Understanding the
limitations of that approach is particularly useful in order to motivate our
choices on how to set up and train neural networks to reproduce this
reweighting. We formulate our NN-based approach in Sec.~\ref{sec:nn} and
describe the toy parton-shower implementation used for validating the approach.
The validation method and its results are presented in
Sec.~\ref{sec:validation}. The approach is then further tested in
Sec.~\ref{sec:example} with a full shower implementation given by the default
\textsc{sherpa} parton shower for the prediction of the Thrust event shape at a
lepton-lepton collider. Finally, we give our conclusions in
Sec.~\ref{sec:conclusions}.

%% file: text/rew_sva.tex
\subsection{On-the-fly reweighting of parton-shower emissions}\label{sec:rsva}

Parton-shower algorithms generate exclusive parton emissions starting from a
simulated event of a given high-energy process. They are based on a
re-formulation of the \textsc{dglap}
evolution~\cite{Gribov:1972ri,Altarelli:1977zs,Dokshitzer:1977sg} using Sudakov
form factors $\Delta$ with an emission kernel $K$~\cite{Sudakov:1954sw}. The Sudakov form
factor gives the probability that no (resolvable) emission occur between two
emission scales $t_\text{low} < t_\text{high}$:
\begin{equation} \label{eq:sudakov}
	\Delta(t_\text{high}, t_\text{low}) = \exp\left(
		-\int_{t_\text{low}}^{t_\text{high}}\text{d}t\, K(t)
	\right)\,.
\end{equation}

We will further specify $t$ and $K$ when we describe our simplified shower
model in Sec.~\ref{sec:shower}. For now, we only need to know how parton-shower
algorithms numerically generate emissions between $t_0$, the starting scale of
the shower usually given by a characteristic scale of the high-energy event
simulated in fixed-order perturbation theory, and
$t_\text{IR}$, the infra-red cut-off scale of the parton shower, where the
evolution would be typically handed over to a fragmentation algorithm to
hadronise the low-energy partons.

The first step is to find the splitting scale $t$ for the next emission. To
achieve this, a random number could be used to sample the kernel $K(t)$ of
Eq.~\eqref{eq:sudakov}. However, doing this in a direct way requires $K$ to be
integrable and invertible. This is not the case for most parton-shower kernels.
A way around this is the Sudakov Veto
Algorithm~\cite{Hoeche:2011fd,Lonnblad:2012hz,Seymour:1994df,
Sjostrand:2006za,Platzer:2011dq,Hoeche:2009xc}.
Herein, one replaces $K$ with a new kernel $\hat{K}$, for which we know the
integral $\hat{\mathcal{K}}$ and its inverse, and which satisfies
$\hat{K}(t)\geq K(t)$ for all $t$. The algorithm then goes as follows:
\begin{enumerate}
	\item
		Set $t\to t_0$, stop if $t_0 < t_\text{IR}$.
	\item \label{sva_item:t_sampling}
		Set $t\to \hat{\mathcal{K}}^{-1}(\log(R_1) + \hat{\mathcal{K}}(t))$
		with a random number $R_1$, stop if $t < t_\text{IR}$.
	\item \label{sva_item:veto}
		Set $P_\text{acc} \to K(t)/\hat{K}(t)$. If $P_\text{acc} > R_2$ for a
		new random number $R_2$, the emission is accepted.
	\item
		Return to Step~\ref{sva_item:t_sampling}.
\end{enumerate}

The hit-or-miss Step \ref{sva_item:veto} counter-balances sampling the
``wrong'' kernel $\hat{K}$ in Step~\ref{sva_item:t_sampling}. When the
algorithm stops, we have a list of scales $t^\text{acc}_i$ for the
generated (i.e.\ accepted) emissions and a list of scales $t^\text{rej}_i$ for
the rejected emissions. We can call this the \emph{parton-shower history} of
the event in terms of $t$.\footnote{For simplicity we show here the sampling
over $t$, but in actual parton-shower algorithms one also samples over two
additional kinematic variables and over the different possible splittings (e.g.
if a gluon splits emits another gluon, or if splits into a quark-antiquark
pair). Only then one has an exclusive shower history.}

Let us ask the following question now: if we have a given shower history
generated according to a kernel $K$, what are the relative probabilities
\begin{equation*}
	w_k = \frac{P(\{t^\text{acc}_i\}, \{t^\text{rej}_i\} | K_k)}
			   {P(\{t^\text{acc}_i\}, \{t^\text{rej}_i\} | K)}
\end{equation*}
to generate the same histories for a set of alternate kernels $K_k$? It turns out
that these probabilities are given by~\cite{Hoeche:2009xc}
\begin{equation}
	\label{eq:rew_fac}
	w_k = \prod_i q^k_\text{acc}(t^\text{acc}_i)
		  \cdot \prod_j q^k_\text{rej}(t^\text{rej}_j)\,,
\end{equation}
where we have defined the reweighting factors
\begin{align*}
	q^k_\text{acc}(t) &= \frac{K_k(t)}{K(t)}\,,\\
	q^k_\text{rej}(t)
		&= \frac{\hat{K}(t) - K_k(t)}{\hat{K}(t) - K(t)}
		 = 1+(1-q_\text{acc}^k(t))\frac{P_\text{acc}(t)}{1-P_\text{acc}(t)}\,.
\end{align*}
The acceptance probability $P_\text{acc}$ in the second form given for
$q^k_\text{rej}$ is defined as in Step~\ref{sva_item:veto} in the Sudakov Veto
Algorithm.

In an event generation, storing the relative probabilities $w_k$ then allows to
reconstruct the spread of an observable under the variations labelled by $k$ in
a more efficient way than if we would do separate re-runs of the simulation for
each $K_k$.
This \emph{parton-shower reweighting} is implemented in the three most commonly
used parton-shower
implementations~\cite{Mrenna:2016sih,Bellm:2016voq,Bothmann:2016nao}. It
complements
reweighting strategies for fixed-order calculations and allows a comprehensive
reweighting of the perturbative parts of an event generation when the interplay
between matrix elements and parton showering is properly accounted for in the
reweighting~\cite{Bothmann:2016nao}.

%% file: text/ps_reweighting.tex
\subsection{The troubles with exact a-posteriori approaches}
\label{sec:psrew}

The applicability of the parton-shower reweighting described in the previous
section is restricted to cases, where the required variations are known
beforehand.  In principle one could store all parton-shower history data needed
to perform reweightings a-posteriori, but as each history will typically
have $\mathcal{O}(10)$ accepted and $\mathcal{O}(100)$ rejected emissions, this
is not practical, in particular because usually the reweighting will not
only depend on the scales $t$, but also on additional kinematic variables of
exclusive emissions, along with the emission channel ($g\to gg$, $g\to
q\bar{q}$, etc.), and possibly the Björken $x$ for initial-state emissions to
be able to evaluate PDF ratios that occur~\cite{Bothmann:2016nao}.

This is a different situation from the one we face when reweighting e.g.\ NLO
and even NNLO calculations, where a much smaller number of values per event is
required to facilitate an exact a-posteriori
reweighting~\cite{Frederix:2011ss,Bern:2013zja,Heinrich:2016jad}.
But even for fixed-order calculations there are
applications for which an event-wise reweighting is not fast enough, as the
number of events is large and possibly tens of thousands of variations are
required. This is the case for instance for PDF fits, where the observables
need to be constantly recomputed along the minimisation process.

This is overcome in the case of fixed-order calculations by averaging over
classes of events that fall into the same observable bin and reweight in the
same way. By the projection onto the observable the individual event kinematics
information can be discarded. The required information can be encoded using an
interpolation grid in a reduced number of kinematic variables (typically the
Björken $x_{1,2}$ and the factorisation scale $\mu_F^2$, such that PDF
variations can be done)~\cite{Carli:2010rw,Kluge:2006xs,Britzger:2012bs,
DelDebbio:2013kxa,Bothmann:2015dba,Bertone:2014zva}.

Can we discard the individual parton-shower history information in a similar
way? For example, we may want to reweight the strong coupling $\alphas$, given
that $K(t)\propto\alphas(t)$. Unfortunately, in the Sudakov Veto
Algorithm, we can not just factorise the ratios $K_k/K$ that occur in
Eqs.~\eqref{eq:rew_fac} and even then the number of thus factorised ratios
would strongly vary. Compare this to fixed-order events, where the dependence
on $\alphas^p$ factorises trivially, and there is only a very limited
set of powers $p$.

Can the combination of interpolation grids and the classification of
parton-shower histories by their similarity with respect to the reweighting
provide a way to reduce the amount of data needed? If we bin the
${t_i^\text{acc}}$ in bins $\beta_\text{acc}$ and
the ${t_i^\text{rej}}$ in bins $\beta_\text{rej}$, we can write an approximation
for Eq.~\eqref{eq:rew_fac} (note that we drop the variation label $k$ for now
to improve readability):
\begin{equation*}
	w^\text{approx}
		= \prod_{\{\beta_\text{acc}\}}
			q_\text{acc}^{w_{\beta_\text{acc}}}(t_{\beta_\text{acc}})
		  \cdot \prod_{\{\beta_\text{rej}\}}
			q_\text{rej}^{w_{\beta_\text{rej}}}(t_{\beta_\text{rej}})\,,
\end{equation*}
where $t_\beta$ is the value of $t$ corresponding to the bin $\beta$, and $w_\beta$ is the
number of emissions that fell into bin $\beta$. We could then jump to the conclusion
that if we can find a way to classify similar parton-shower histories into
classes $c$ that behave similarly under variations, we could write
\begin{equation} \label{eq:class_rew}
	\langle w^\text{approx} \rangle_c
		= \prod_{\{\beta_\text{acc}\}}
			q_\text{acc}^{\langle w_{\beta_\text{acc}} \rangle_c}(t_{\beta_\text{acc}})
		  \cdot \prod_{\{\beta_\text{rej}\}}
			q_\text{rej}^{\langle w_{\beta_\text{rej}} \rangle_c}(t_{\beta_\text{rej}})\,.
\end{equation}
Suppose we calculate the $\langle w_\beta \rangle_c$ for every $t$-histogram bin $\beta$ during
a pre-production run, and the relative proportions of events $r_c$ that feature
a parton-shower history that is classified into $c$. We could then calculate
the effect of the parton shower variation by replacing the nominal cross
section~$\sigma$ with
\begin{equation*}
	\sigma \to \sigma  \cdot \sum_c r_c \langle w^\text{approx} \rangle_c\,.
\end{equation*}
However, this ansatz has a critical flaw, which is the nature of the average on
the left-hand side of Eq.~\eqref{eq:class_rew}. The arithmetic means on the
right-hand side are in the exponent of the reweighting functions, and thus the
left-hand side mean is a \emph{geometric mean}. It follows that
\begin{equation*}
	\langle w^\text{approx} \rangle_c
		=    \langle w^\text{approx} \rangle_c^\text{geometric}
		\leq \langle w^\text{approx} \rangle_c^\text{arithmetic}\,.
\end{equation*}
Unfortunately we need to know the right-hand side of this last equation, but
for that we need to retain the individual shower history information. And hence
this classification ansatz to discard that information fails.

Lacking a straightforward analytical way to solve the problem, we will therefore
in the following present a proof-of-principle for using a simple neural network
to find the bin-wise reweighting factors needed to vary the parton shower. The
problem can be formulated in the following way. Let us assume that we want to
reweight the cross section of an observable bin $b$ as we vary a set of
parameters that we will collectively denote $\theta$.
For each event generated in the Monte Carlo sample,
which we label with an index $i$, there is a reweighting factor $w_{b,i}$. As
shown in Eq.~\eqref{eq:rew_fac}, the event-wise reweighting factor $w_{b,i}$ is
a functional of $q_\text{acc}(t;\delta\theta)$, where we have written explicitly
the dependence of the latter on the variation of the parameters, $\delta\theta$.
Different variations of the parameters yield different functions
$q_\text{acc}(t;\delta\theta)$, and hence different reweightings.

The key assumption underlying this study is that we can ignore the details of
the shower history, and therefore the analytical expression for $w_{b,i}$.
Instead we model the dependence of the average reweighting factor for a given
bin, $\langle w_b\rangle$, on the function $q_\text{acc}(t)$ using a neural
network. After training on a discrete set of variations
$\{\delta\theta^{(k)}\}$ labelled by $k\in \mathcal{T}$, we expect the NN to be
capable of interpolating to the correct value of the reweighting factor for a
generic variation. All dependencies should be sufficiently smooth that the NN
can produce a satisfactory interpolation. The output of the NN can be validated
against correct reweighting factors before using it in an application.


%% file: text/shower_model.tex
\subsection{Simplified shower model} \label{sec:shower}

To formulate and validate our approach it is sufficient to use a simplified
parton-shower model. First, let us further specify the kernel we use in the
Sudakov form factor:
\begin{equation} \label{eq:kernel}
	K(t)
	= \int_{\varepsilon(t)}^{1-\varepsilon(t)} \text{d}z\, K(t, z)
	= \int_{\varepsilon(t)}^{1-\varepsilon(t)} \text{d}z\, \frac{1}{t}
	  \frac{\alphas(\mu^2(t))}{2\pi}
	  \sum_{a\to bc}
	  P_{ab}(z)\,.
\end{equation}
The variable $z$ gives the energy ratio between the mother parton $a$ and its
daughter $b$. We define $t$ to be such that $z(z-1)t$ is proportional to the
transverse momentum squared of the emitted particle, $p_{T,b}^2$ (with respect
to $a$). Beyond that, the precise definition of $t$ and the emission kinematics
are not important for our purposes. However, together with the requirement
$t>t_\text{IR}$ this allows us to specify what a resolvable emission is, by
setting the integration limits for $z$ using $\varepsilon(t) = t_\text{IR}/t$.
Thus, the kinematic limits on $z$ ensure that the transverse momenta of all
generated emissions is larger than $t_\text{IR}$. This also takes care of the
singularities that appear in some of the \textsc{dglap} splitting functions
$P_{ab}=P_{a\to bc}$ at $z=0$ and $z=1$. In our simple shower, we only use the
following three LO \textsc{dglap} splitting functions as they are listed
in~\cite{Buckley:2011ms}:
\begin{align*}
	P_{q\to qg}(z) &= C_F \, \frac{1 + z^2}{1 - z}\,, \\
	P_{g\to gg}(z) &= C_A \, \frac{z^4 + 1 + (1 - z)^4}{z (1 - z)}\,, \\
	P_{q\to qg}(z) &= T_R \, n_f \, \left(z^2 + (1 - z)^2\right)\,,
\end{align*}
where $C_F = 4/3$, $C_A=3$, $T_R=1/2$ and $n_f=5$.

It remains to define the scale $\mu^2(t)$ at which the strong coupling
$\alphas$ is evaluated. In the most simple shower model, one could just
set it to $\mu^2(t)=t$. However, we do not want to move the needle too far in
the direction of a simplified shower model. Instead, we keep some of the
complications of currently used shower algorithms. Therefore, we set the scale
to be $\mu^2(t)=z(z-1)t$. This choice implicitly includes higher-order
corrections in Eq.~\eqref{eq:kernel}~\cite{Catani:1990rr}.

We intend to sample over the now explicit $z$ dependence and over the splitting
channel $a\to bc$ (and over the different $a$ in the parton cascade). This
introduces adjustments to the Sudakov Veto Algorithm as it is described
in~\ref{sec:rsva}, cf.~\cite{Lonnblad:2012hz}.  For the reweighting, we use the
integrands where $z$ is not integrated out:
\begin{align*}
	w_k &= \prod_i q_\text{acc}^k(t^\text{acc}_i, z^\text{acc}_i)
		   \cdot \prod_j q_\text{rej}^k(t^\text{rej}_j, z^\text{rej}_j)\,,\\
	q_\text{acc}^k(t, z)
		&= \frac{K_k(t, z)}{K(t, z)}\,,\\
	q_\text{rej}^k(t, z)
		&= 1+(1-q_\text{acc}^k(t, z))\frac{P_\text{acc}(t, z)}{1-P_\text{acc}(t, z)}\,,
\end{align*}
and therefore we need to know $(t, z)$ for every accepted or rejected splitting
for the reweighting. In the following we will only consider reweighting as we
vary the values of the strong coupling $\alphas$. Since everything else in the
ratio $K_k/K$ cancels, the reweighting functions simplify:
\begin{subequations}
\label{eq:rew}
\begin{align}
	w_k &= \prod_i q_\text{acc}^k(\mu^2_{\text{acc},i})
		   \cdot \prod_j q_\text{rej}^k(\mu^2_{\text{rej},j}, P_{\text{acc},j})\,,\\
	\label{eq:qacc}
	q_\text{acc}^k(\mu^2)
		&= \frac{\alpha_{\text{S},k}(\mu^2)}{\alphas(\mu^2)}\,,\\
	\label{eq:qrej}
	q_\text{rej}^k(\mu^2, P_\text{acc})
		&= 1+(1-q^k_\text{acc}(\mu^2))\frac{P_\text{acc}}{1-P_\text{acc}}\,,
\end{align}
\end{subequations}
and we only need to know $\mu^2$ for accepted emissions, and $(\mu^2,
P_\text{acc})$ for rejected emissions.

To conclude the definition of our shower model, it remains to define the
initial and cut-off conditions for the shower evolution. The starting scale
$t_0$ would be usually determined by the high-energy event. However, we do not
simulate that and instead randomly sample $t_0$ for each shower history from a
Gaussian distribution with a mean value and standard deviation of
$\SI{e4}{\GeV\squared}$. The starting configuration is that of a single
(final-state) quark line. The infra-red cut-off is chosen to be $t_\text{IR} =
\SI{1}{\GeV\squared}$.


%% file: text/nn_io.tex
\subsection{Input and output data}
\label{sec:nn_io}

As discussed at the end of Sect.~\ref{sec:psrew} our goal is to replace the
reweighting of single parton-shower histories as described in
Eqs.~\eqref{eq:rew} with an a-posteriori reweighting of individual histogram
bins after filling them with showered events.  So, for a given variation of the
parameters, $\delta\theta^{(k)}$, we need to predict the average over reweighting
factors $w_{k,i}$ for histories of events $i$ that fell into a given observable
bin $b$. Let us designate this quantity as $\langle w_k \rangle_b$. Then, in
the limit of infinite statistics, we know that if with the nominal calculation
we find that $N_b$ events fell into the bin $b$, then for the variation $k$ we
will find $\langle w_k \rangle_b N_b$ events.

So if we use a neural net for this purpose, it is clear that $\langle w_k
\rangle_b$ should be the value of our single output neuron, and we can train
the neural net against this value (which can be obtained from an on-the-fly
reweighting for the variation $k$, or from a separate re-run for this
variation).

The input data is more ambiguous. It has to be a trade-off between precisely
describing the variation and using as few data points as possible. Our input is
a real vector of size $N_\text{in}$ obtained by sampling the acceptance function
$q_\text{acc}^k(\mu^2)$ defined in Eq.~\eqref{eq:qacc} at a set of discrete points
$\mu^2_\ell$, where $\ell=1,\ldots,N_\text{in}$. We use a logarithmic
distribution of $\mu^2$ points for the sampling (i.e.\ with more points for lower
scales), because the $\alphas$ ratios we will be using change more quickly for
lower scales. Note that $q_\text{acc}^k$ also appears in the reweighting function
for rejected events $q_\text{rej}^k$, cf. Eq.~\eqref{eq:qrej}, and therefore at
least partly specifies how to vary rejected emissions.

\begin{figure}
	\begin{subfigure}[t]{0.49\textwidth}
		\includegraphics{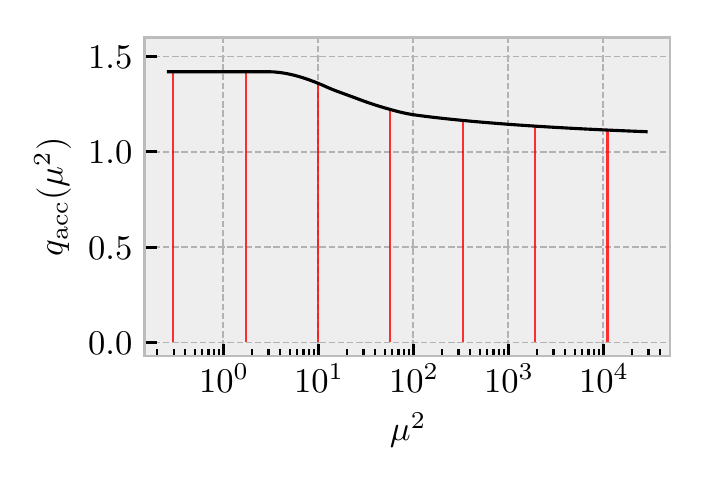}
		\caption{Neural net input data example}
	\end{subfigure}
	\hfill
	\begin{subfigure}[t]{0.49\textwidth}
		\includegraphics{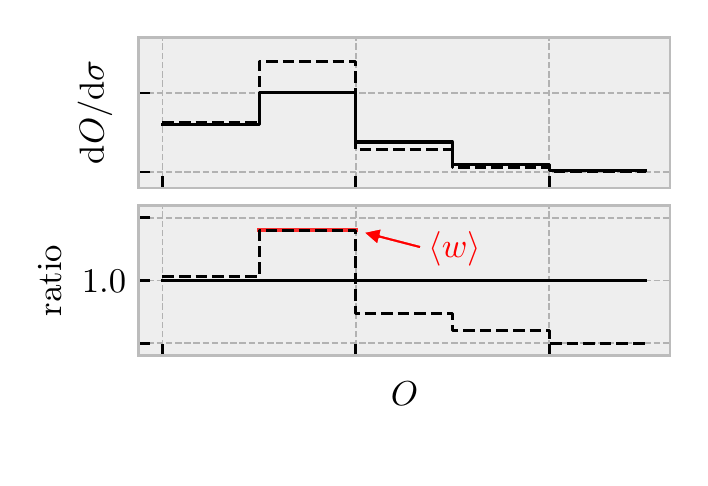}
		\caption{Corresponding neural net output data}
	\end{subfigure}
	\caption{
		Visualisations of the neural network input/output data for a given
		observable bin (here the second bin in the histogram for an observable
		$O$, depicted in the right-hand panel). Given a number of values of
		$q_\text{acc}$ for a variation (left-hand panel) as input, the NN is
		trained such that it outputs the corresponding value $\langle w
		\rangle$ for the given bin.
		}
	\label{fig:nn_io_visualisation}
\end{figure}

For illustration purposes, we show visualisations of these input/output choices
in Fig.~\ref{fig:nn_io_visualisation}. The panel on the left shows the function
$q_\text{acc}$, the red lines correspond to the values of $\mu^2$ at which we
sample the function for the input vector. The shift in the observable in each
bin is reported in the panel on the right, where the solid line shows the value
of the observable for the nominal value of the parameters $\theta$, and the
dotted line shows the shift that is observed for a variation of such
parameters. The NN output needs to reproduce the shift in the ratio of the
nominal value and the variation value.

%% file: text/nn_arch_and_training.tex
\subsection{Neural-network architecture and training}
\label{sec:nnarch}

For our neural-network implementation and training we we use the
\textsc{PyTorch} python library~\cite{Paszke:2017aut}.

The input data is fed into a layer that consists of $N_\text{in}=60$ neurons
that are \emph{linear modules}, i.e. they compute their output using a linear
function $y=mx+c$ from the input $x$, with the weight $m$ and the bias $c$ being
trainable variables. As we have discussed in the previous section, the input $x$
is given by one of $N_\text{in}$ function values of the $q_\text{acc}^k$ function.

The input layer is fully connected to the next layer, which consists of 15
neurons that are \emph{rectifier linear units} (\textsc{ReLU}). Given a value
$x$ from
the input layer, they pass on the value $y=\max(0,x)$ to the output layer. It
is this hidden layer that introduces a non-linearity to the network, which is
surely required for the problem at hand, given Eqs.~\eqref{eq:rew}.

The \textsc{ReLU} layer is then fully connected to the output layer which is
just a
single neuron. It is a linear module like the ones in the input layer. Its
output $y$ gives the neural-network prediction for $\langle w_k \rangle_b$,
i.e. $y=\langle w_k \rangle_b^\text{NN}$.

The training is done by minimizing the squared Euclidean distance between
$\langle w_k \rangle_b^\text{NN}$ and the training data $\langle w_k
\rangle_b$ for a range of different variations $k \in \mathcal{T}$,
where $\mathcal{T}$ specifies the training data set.
The {\em loss function} is defined as
\begin{equation}
    \mathcal{L} = \sum_{k\in\mathcal{T}}
    \left[
        \langle w^{(k)} \rangle_b^\text{NN} - \langle w^{(k)}
        \rangle_b
    \right]^2\,.
\end{equation}
The optimisation step is performed using the \textsc{Adam}
algorithm~\cite{Kingma:2014aa} as implemented in \textsc{PyTorch}, with a
learning rate of $10^{-4}$. The learning passes are performed until either a
preset optimisation target (Euclidean distance $\leq 10^{-5}$) or a maximum
number of passes ($10^6$) is reached. On a \SI{2.8}{GHz} Core i7 processor
this caps the CPU time for a training at approximately \SI{30}{\second},
although depending on the training data and on the random initialization,
it can also take only a few seconds. If necessary, the training time could be
further reduced by using the GPU acceleration supported by \textsc{PyTorch}.

Every 5000 optimisation steps it is checked if the loss function has reduced by
at least \SI{0.1}{\percent}. If this is not the case for five subsequent
checks, the training is cancelled, assuming that a stable minimum has been
found. If this minimum is however associated with a loss function value that is
greater than \SI{1}{\percent}, the whole training pass is discarded and
repeated with a newly randomised set of NN weights. Enforcing this
threshold ensures that we do not accept trainings that have become stuck in a
non-global minimum.

%% file: text/validation_procedure.tex
\subsection{Validation procedure and training data}

In this section we compare neural-network predictions $\langle w_k
\rangle_b^\text{NN}$ with the true reweighting factor $\langle w_k \rangle_b$
for a range of different variations $k \in \mathcal{T}$ and observable bins $b$
(which are defined below). In doing that we will always exclude the data point
that we want to predict from the trainings.

First, we generate $10^6$ parton-shower histories with the simplified model
described in Sec.~\ref{sec:shower} and bin them one-by-one into a bin $b$ for
each observable. For each observable bin $b$ and variation $k$, we calculate
$\langle w_k \rangle_b$. These values will be the  output data sample for
the training/comparison of our neural networks.

As the input data sample, we calculate $N_\text{in}=60$ function values for
$q_\text{acc}^k(\mu^2)$ for each variation $k \in \mathcal{T}$,
distributed logarithmically
between the lowest scale $\min(\mu^2)=t_\text{IR}/4$ reachable by the shower and
$\mu_{t_0}+2\sigma_{t_0}$, i.e. including large $\mu^2$ values up two standard
deviations $\sigma_{t_0}$ from the mean of the starting scale distribution
$\mu_{t_0}$.

The general training procedure described in Sec.~\ref{sec:nnarch} is amended for the
validation as follows. For predicting $\langle w_{k'} \rangle_{b'}$ for a given
variation $k'$ and histogram bin $b'$, we use all the input and output data for
$b'$ and all $k \in \mathcal{T}$ \emph{except for} $k=k'$. Note that this means
that we repeat the training for each $k \in \mathcal{T}$ (because we do not
want to include the data for $k'$ in the validation). In an application on the
other hand, only one neural network for each observable bin $b$ would need to
be trained to interpolate between the $k$.

Lastly, we repeat this $N_\text{trainings}=10$ times, each with randomly
initialised neural net weights. Our neural net prediction for $(k',b')$ is then
given by the mean of the outputs of these 10 neural nets given the values for
$q_\text{acc}^{k'}$ as input. As an uncertainty, we give the standard deviation
for these 10 values.

\begin{figure}
	\centering
	\includegraphics{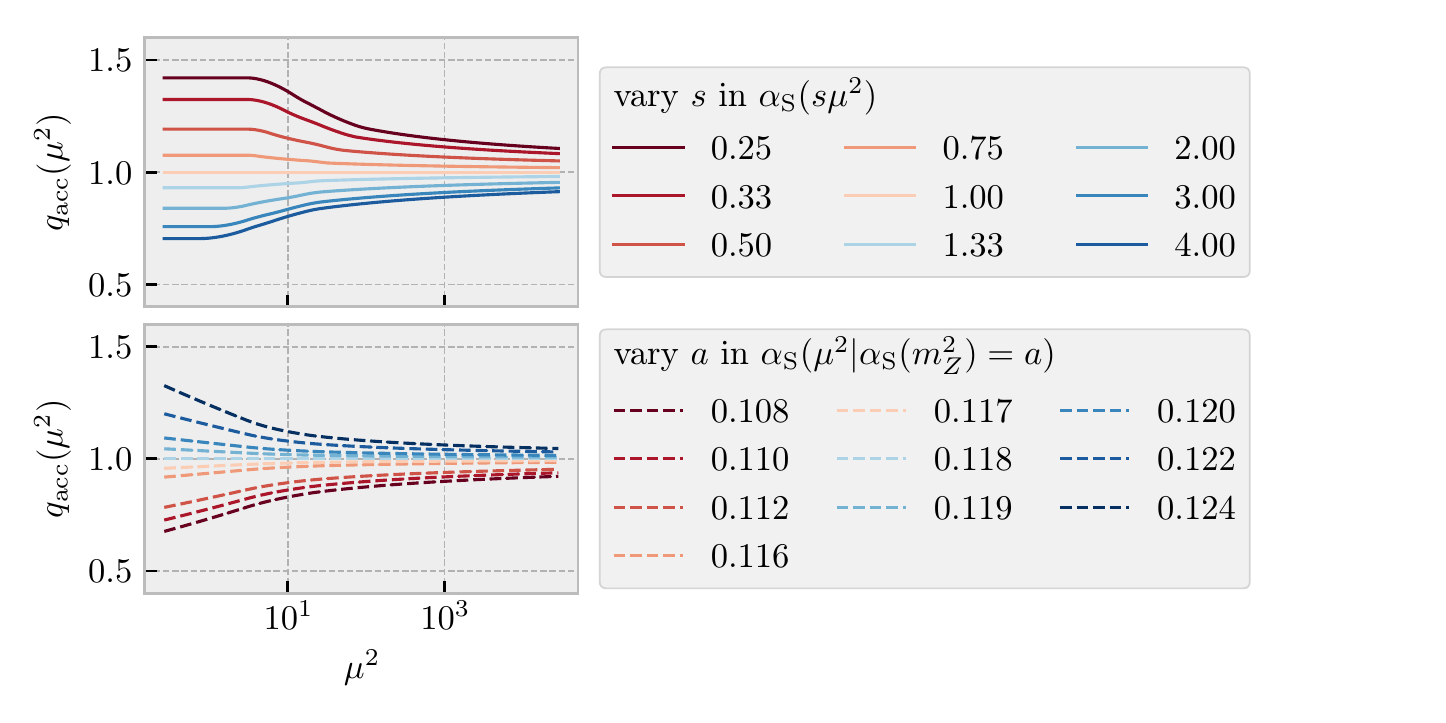}
	\caption{
		Display of the $\alphas$ variations used for validating the
		neural network approach. The upper row shows the scale
		variations, whereas the lower row shows variations of
		the input value $\alphas(m_Z^2)$. The $q_\text{acc}$ function
		values as shown on the left are the input for the neural nets, as
		explained in Sec.~\ref{sec:nn_io}.
		}
	\label{fig:qacc}
\end{figure}

Now let us define the training data set $\mathcal{T}$. The variations used in
the training should be diverse enough to allow the neural net to predict other
variations accurately. We include two classes of variations of $\alphas$:
\begin{alignat}{2}
	\alpha_\text{S}(\mu^2) &\rightarrow
		\alphas(s\,\mu^2)\,,
	\qquad& s &= 0.25\dots4.00\,,
	\\
	\alpha_\text{S}(\mu^2) &\rightarrow
		\alphas^{}(\mu^2\,|\,\alphas(m_Z^2)=a)\,,
	\qquad& a &= 0.108\dots0.124\,,
\end{alignat}
where $\alphas^{}(\mu^2\,|\,\alphas(m_Z^2)=a)$ is defined to be the strong
coupling value at $\mu^2$ given that the value at the $Z$-boson mass,
$\alphas(m_Z^2)$, is set to $a$. For convenience we retrieve all $\alphas$
values we use from \textsc{nnpdf}\,3.1 sets~\cite{Ball:2017nwa} interfaced
using \textsc{lhapdf}~\cite{Buckley:2014ana}. This also dictates our choice of
$a=\alphas(m_Z)$ variations, since we use the values available in that
\textsc{nnpdf} release. The individual values of $s$ and $a$ used in the
validation and the corresponding $q_\text{acc}$ functions are shown in
Fig.~\ref{fig:qacc}.

%% file: text/results.tex
\subsection{Results}

We discuss the results for the validation strategy laid down in this section so
far for histogram bins of two observables: the scale $t^\text{lead}$ of the
first emission in a parton-shower history, and the number of accepted emissions
$N_\text{em}$ in a history. Note that these observables are not physical (as we
use no jet algorithm), but
merely serve here to test our approach.

\subsubsection{Leading emission scale}

\begin{figure}
	\centering
	\includegraphics{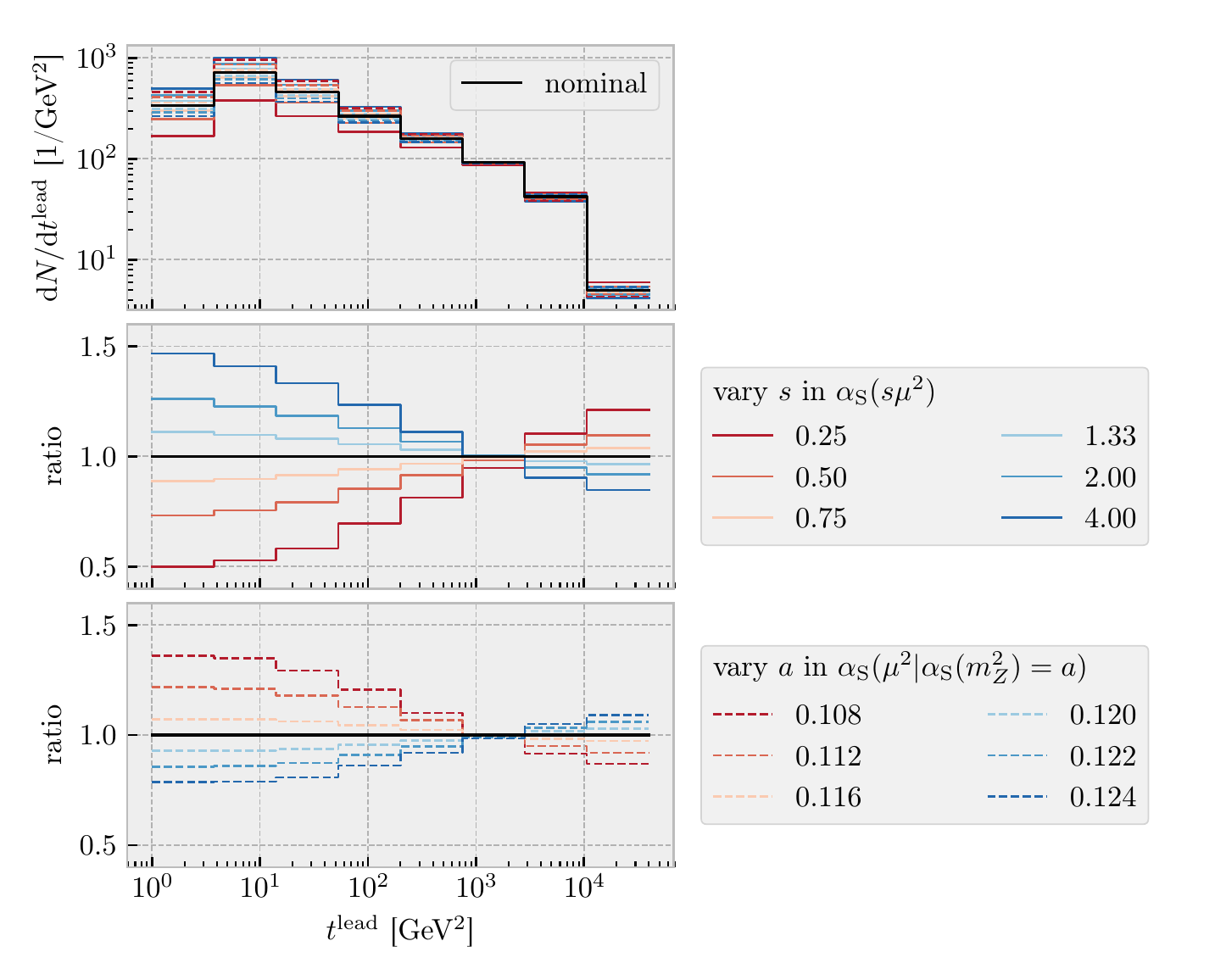}
	\caption{
		The effect of parton-shower \alphas\ variations on the distribution of
		the leading emission scale $t^\text{lead}$. The lower two
		panels show the ratios broken down into
		scale variations and $\alphas(m_Z^2)$ variations, respectively.
		Some intermediate variations that are used for the validation are left
		out here for clarity.
		The black line (``nominal'') gives
		the distribution for the nominal \alphas\ choice, i.e.\ for $s=1$ and
		$a=0.118$.
		}
	\label{fig:ptlead_distribution}
\end{figure}

\begin{figure}
	\centering
	\includegraphics{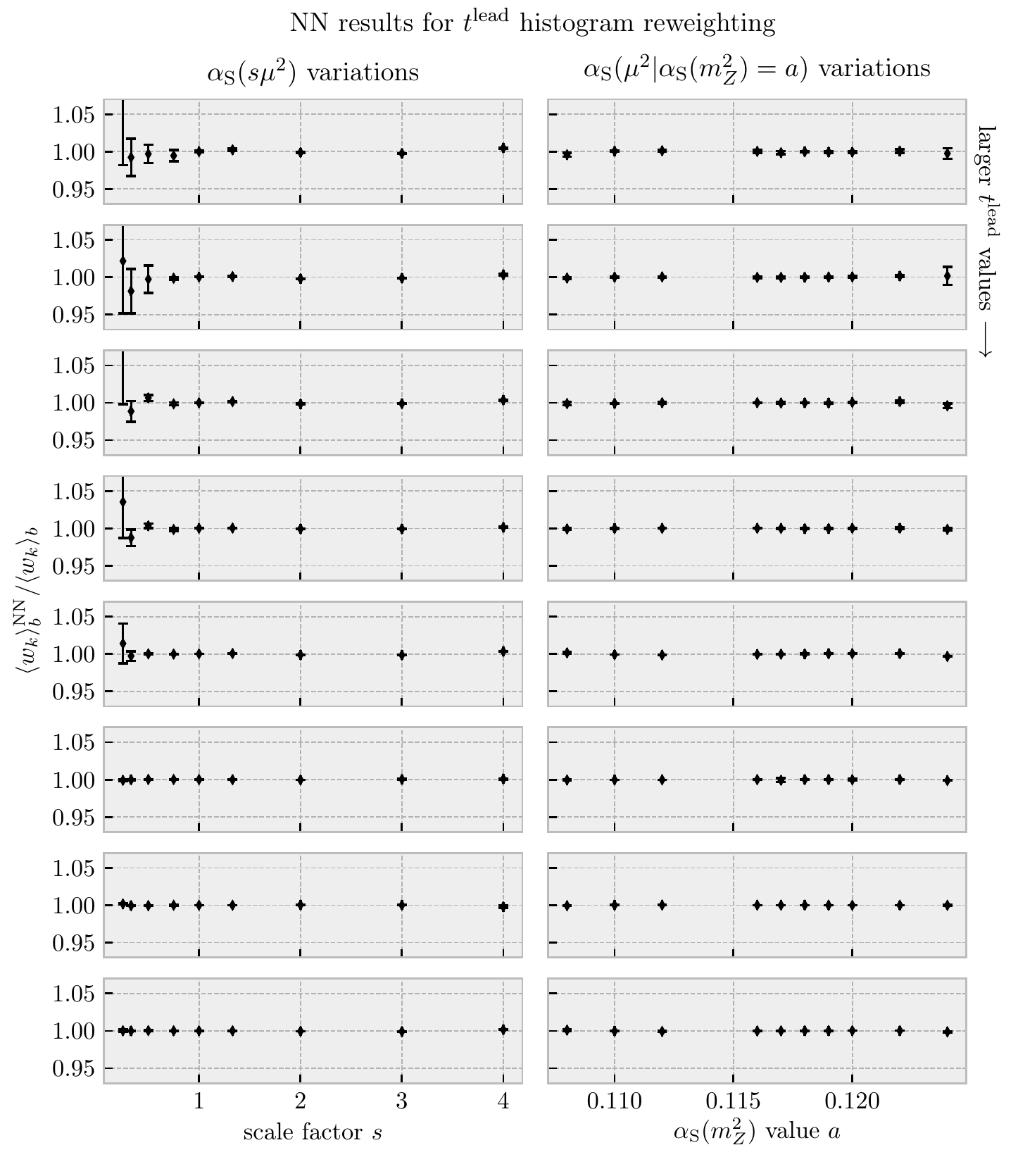}
	\caption{
		Ratios between neural-net-predicted and true reweighting factors for
		each bin of the $t^\text{lead}$ histogram, see
		Fig.~\ref{fig:ptlead_distribution}. Each row corresponds to one bin,
		with the smallest-$t^\text{lead}$ bin at the top. The two columns
		correspond to scale and $\alphas(m_Z^2)$ variations.
		}
	\label{fig:ptlead_results}
\end{figure}

The first observable is the scale of the hardest emissions $t^\text{lead}$,
binned into a histogram of 8 logarithmically distributed bins between 1 and $3
\cdot 10^4$ GeV$^2$. The $t^\text{lead}$ distribution for the nominal $\alphas$
values and a subset of variations is shown in
Fig.~\ref{fig:ptlead_distribution}. The ratios between the NN result for the
reweighting factor $\langle w_k \rangle_b^\text{NN}$ and the true value
$\langle w_k \rangle_b$ are shown in Fig.~\ref{fig:ptlead_results}, for all
variations $k$ and observable bins $b$.

For the interpretation of this figure, note that we actually display ratios of
ratios, i.e.\ the deviation between reweighting factors, that are itself just
ratios
of the nominal and the varied cross section. To calculate how far we are off
relative to the absolute value of the cross section, one would need to multiply
the true reweighting factor with the ratio in Fig.~\ref{fig:ptlead_results}.

It is clear from the plots that the NN results are within
\SI{1}{\percent} of the true reweighting factor, and the uncertainty of the
prediction is $\pm\SI{3}{\percent}$ or less. Exceptions only appear for the
$s=0.25$ variation at small values of $t^\text{lead}$. Note that $s=0.25$ is an
extremal variation, and hence in our validation procedure (where the
to-be-predicted point is not included in the training data set) the NN
needs to \emph{extrapolate} when predicting this reweighting factor. The
training data set should therefore go beyond the variations that are to be
expected in an application, such that all NN predictions are guaranteed to be
interpolations.

We have also tested how the uncertainties of the prediction behave when we
initialise the NN weights with the same random numbers for each
repetition of the training procedure. In that case, the spread becomes
negligible compared to the results in Fig.~\ref{fig:ptlead_results}. This means
that the spread is not due to the numerical precision (floating point
precision), but due to the random initialisation. Hence, in order to get even
more precise predictions, the NN architecture and/or training procedure
would have to be modified.

\subsubsection{Number of emissions}

\begin{figure}
	\centering
	\includegraphics{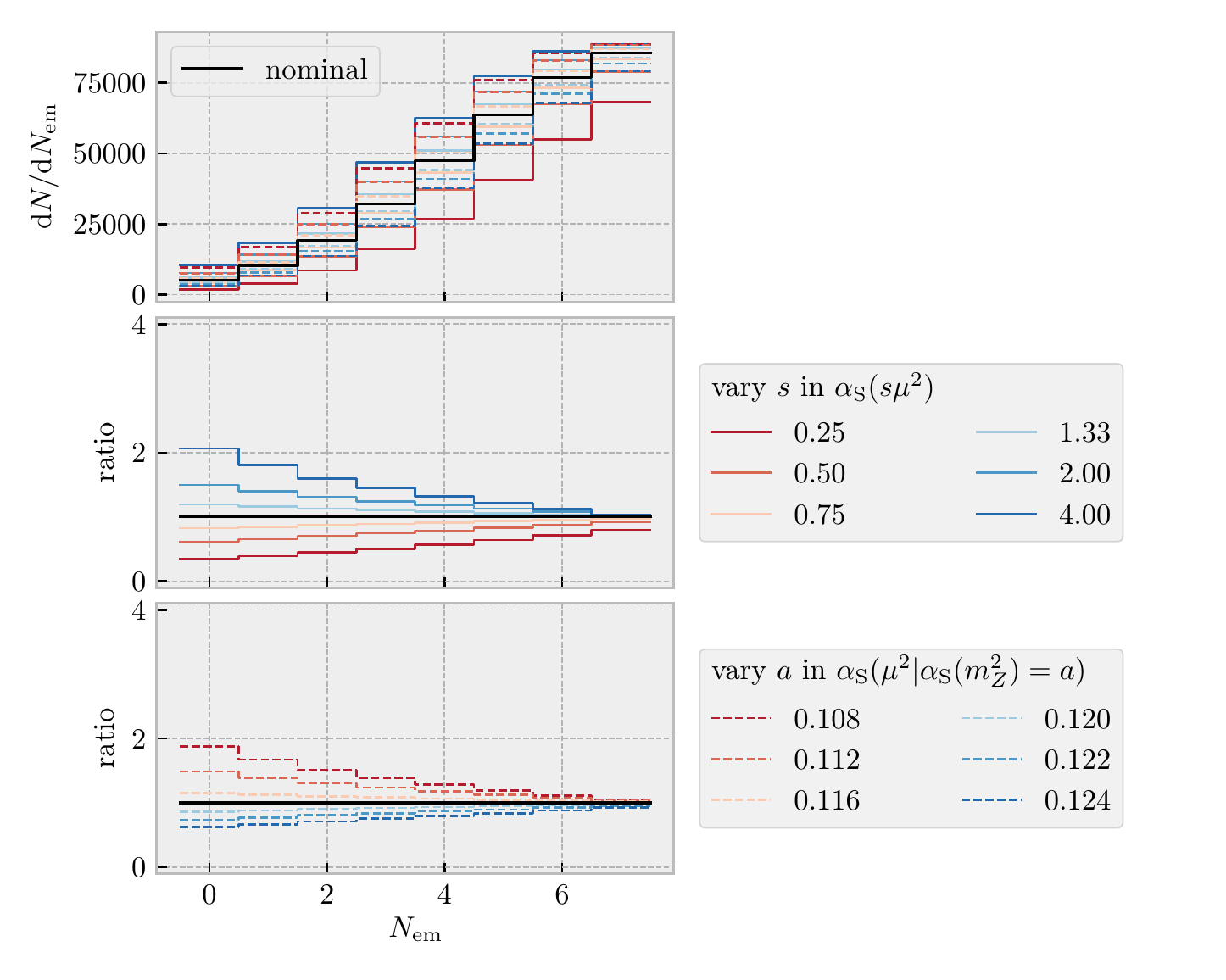}
	\caption{
		The effect of parton-shower \alphas\ variations on the distribution of
		the number of emissions $N_\text{em}$. The lower
		two panels show the ratios broken down into scale variations and
		$\alphas(m_Z^2)$ variations, respectively. See
		Fig.~\ref{fig:ptlead_distribution} for additional notes.
		}
	\label{fig:nemissions_distribution}
\end{figure}

\begin{figure}
	\centering
	\includegraphics{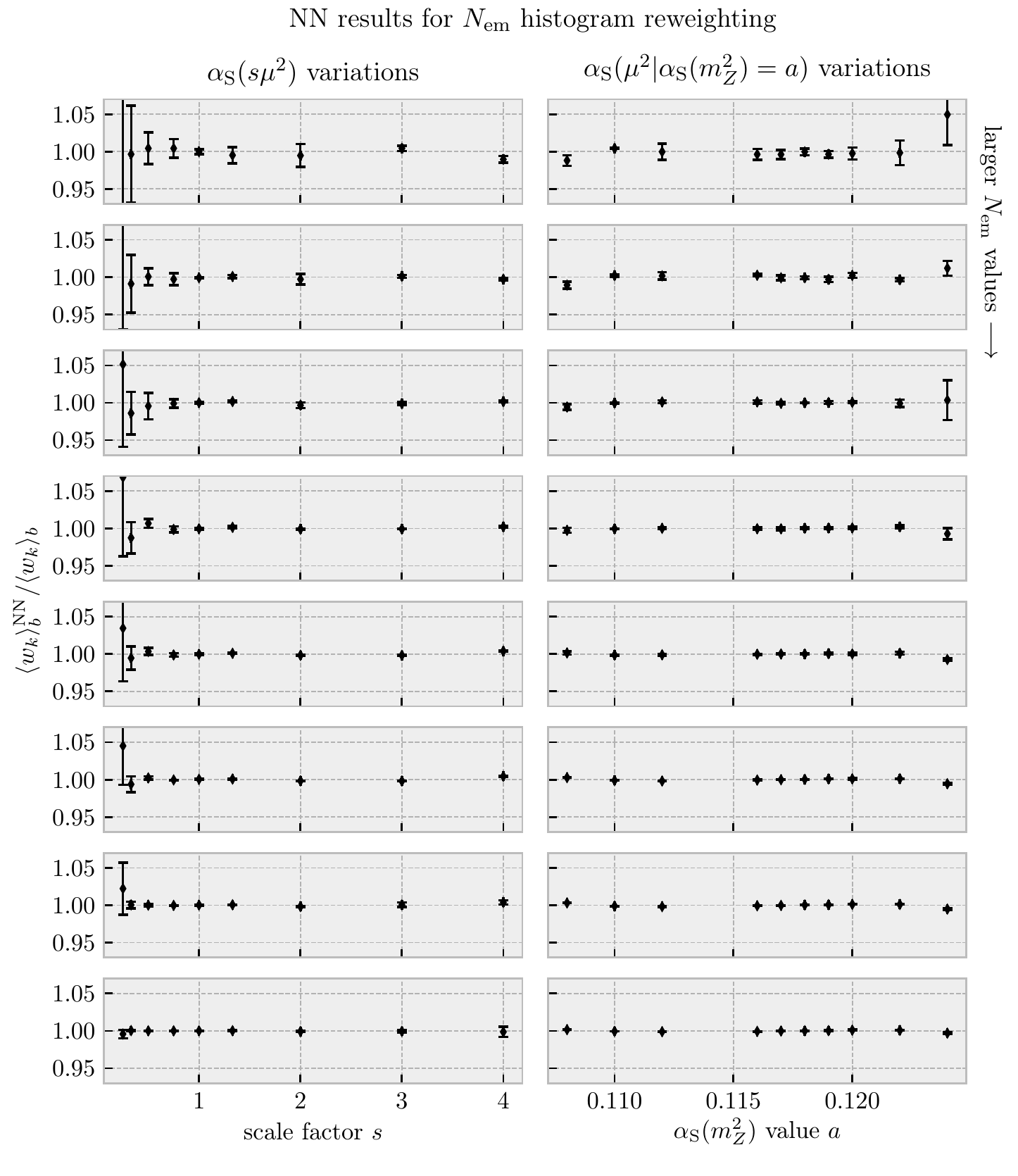}
	\caption{
		Ratios between neural-net-predicted and true reweighting factors for
		each bin of the $N_\text{em}$ histogram, see
		Fig.~\ref{fig:nemissions_distribution}. Each row corresponds to one
		bin, with the smallest-$N_\text{em}$
		bin at the top. The two columns
		correspond to scale and $\alphas(m_Z^2)$ variations.
		}
	\label{fig:nemissions_results}
\end{figure}

We also test our approach with the number of emissions, in 8 bins between 0
and 7~emissions. As for the $t^\text{lead}$, we show the $N_\text{em}$ histogram
with all variations in Fig.~\ref{fig:nemissions_distribution}, and the ratios
for the reweighting factors given by the neural network and the training data
in Fig.~\ref{fig:nemissions_results}.

As for the $t^\text{lead}$ case we find that the neural network predictions are
within \SI{1}{\percent} of the true reweighting factor, with an uncertainty
that only for some cases exceeds \SI{3}{\percent}. The exceptions occur in the
region for variations that enhance emission probabilities (small $s$ and/or
large $a$), in particular for bins with smaller $N_\text{em}$ values.

\subsubsection{Further tests}

\begin{figure}
	\centering
	\begin{subfigure}[t]{\textwidth}
		\includegraphics{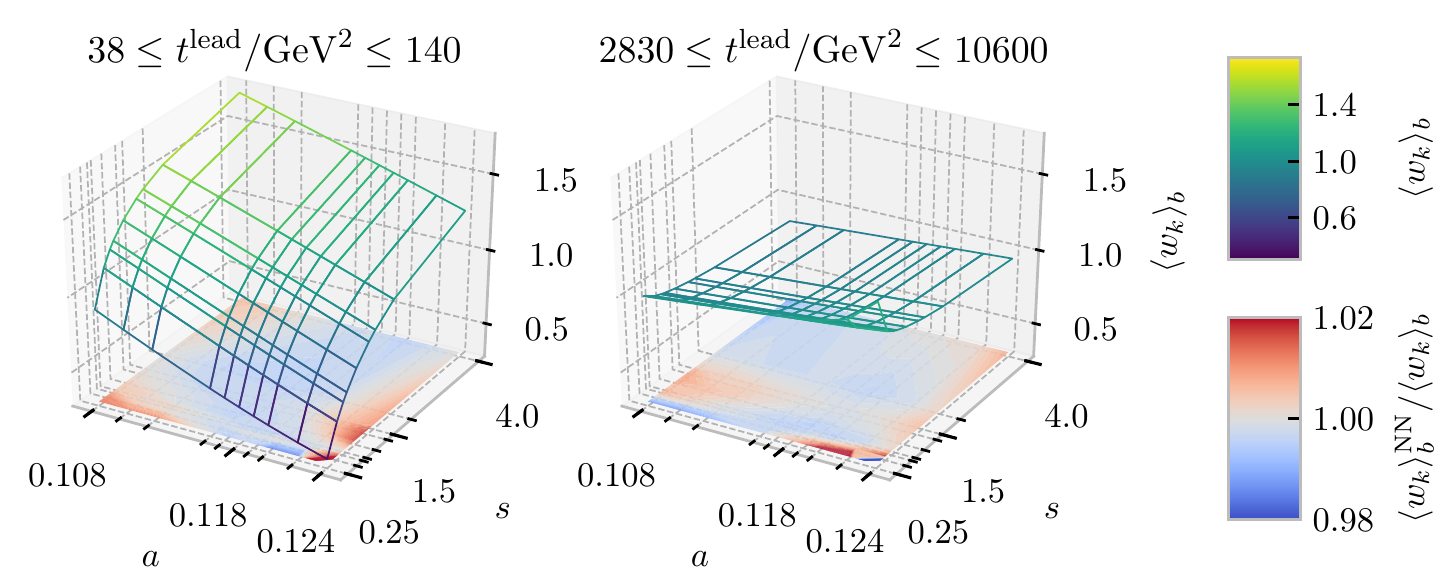}
		\caption{\label{fig:3d_results:ptlead}
			leading emission scale $t^\text{lead}$
			}
	\end{subfigure}
	\begin{subfigure}[t]{\textwidth}
		\includegraphics{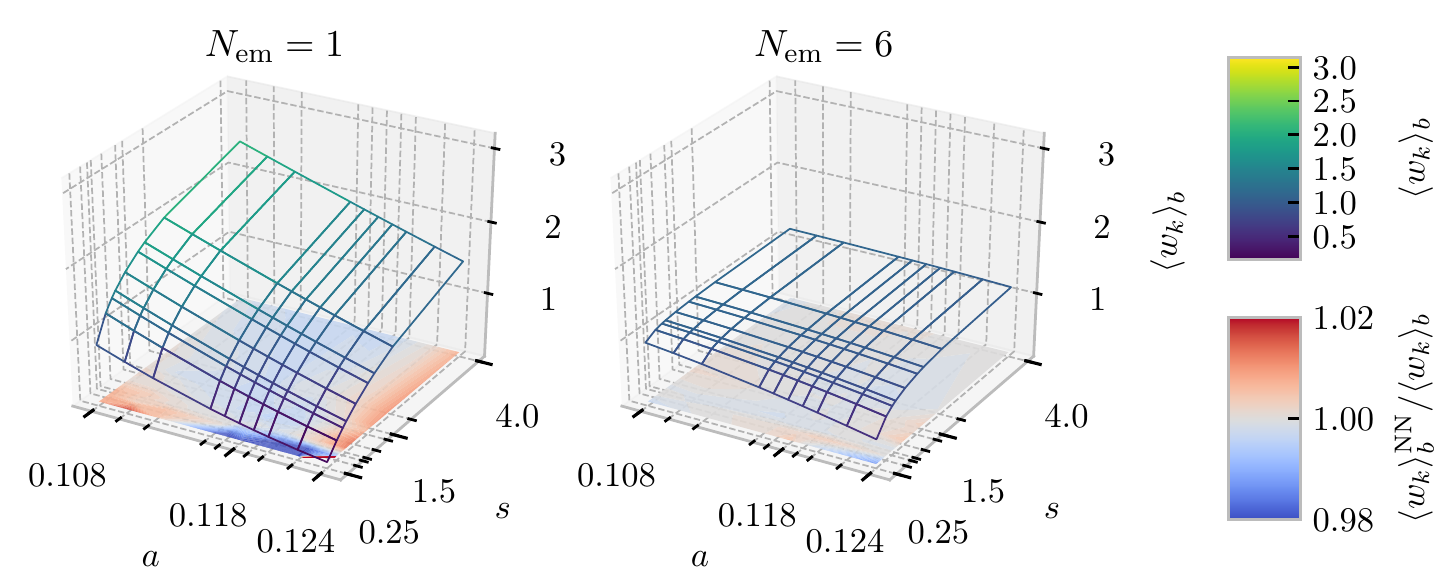}
		\caption{\label{fig:3d_results:nemissions}
			number of emissions $N_\text{em}$
			}
	\end{subfigure}
	\caption{
		The reweighting factors for a low- and a high-$t^\text{lead}$
		($N_\text{em}$) bin are shown in the upper (lower) row. The projections
		on the $a$-$s$ plane show the ratio between neural-net-predicted and
		true reweighting factors, $\langle w_k \rangle_b^\text{NN}/\langle
		w_k \rangle_b$. The predicted factor was omitted in the training of
		the corresponding neural net.
		The clipped corners in the ratio projection for the
		$t^\text{lead}$ bins and the $N_\text{em}=1$ bin are due to the ratio
		being beyond the scale of $\pm \SI{2}{\percent}$ for $a=0.124$ and
		$s=0.25$.
		}
	\label{fig:3d_results}
\end{figure}

In Fig.~\ref{fig:3d_results:ptlead} (\ref{fig:3d_results:nemissions}) we
present the absolute reweighting factors for a low- and a high-$t^\text{lead}$
($N_\text{em}$) bin. Here we also include simultaneous variations (i.e.\ both
$s \neq 1.0$ \emph{and} $a \neq 0.118$),
the training data set is defined by all pairs $(a, s) \in A \otimes S$ with the
$\alphas(m_Z^2)$ values
$A=\{0.108,0.110,0.112,0.116,0.117,0.118,0.119,0.120,0.122,0.124\}$ and the
scale factors $S=\{0.25, 0.5, 1.0, 1.5, 2.0, 4.0\}$.
The ratio of the NN result over
the true reweighting factor is shown as a projection below the absolute
reweighting factors. To calculate this, we use the same validation method as
before,
i.e.\ the predicted reweighting factor is left out in the training of the
NN that is used for this point. We find that the deviations are within
\SI{2}{\percent} for simultaneous variations, except for the variation
$a=0.124$, $s=0.25$, for which emission probabilities are maximally enhanced
and $q_\text{acc}$ is most non-linear.
This findings are also true for the bins that are not shown in the figure.

\begin{figure}
	\centering
	\begin{subfigure}[t]{\textwidth}
		\includegraphics{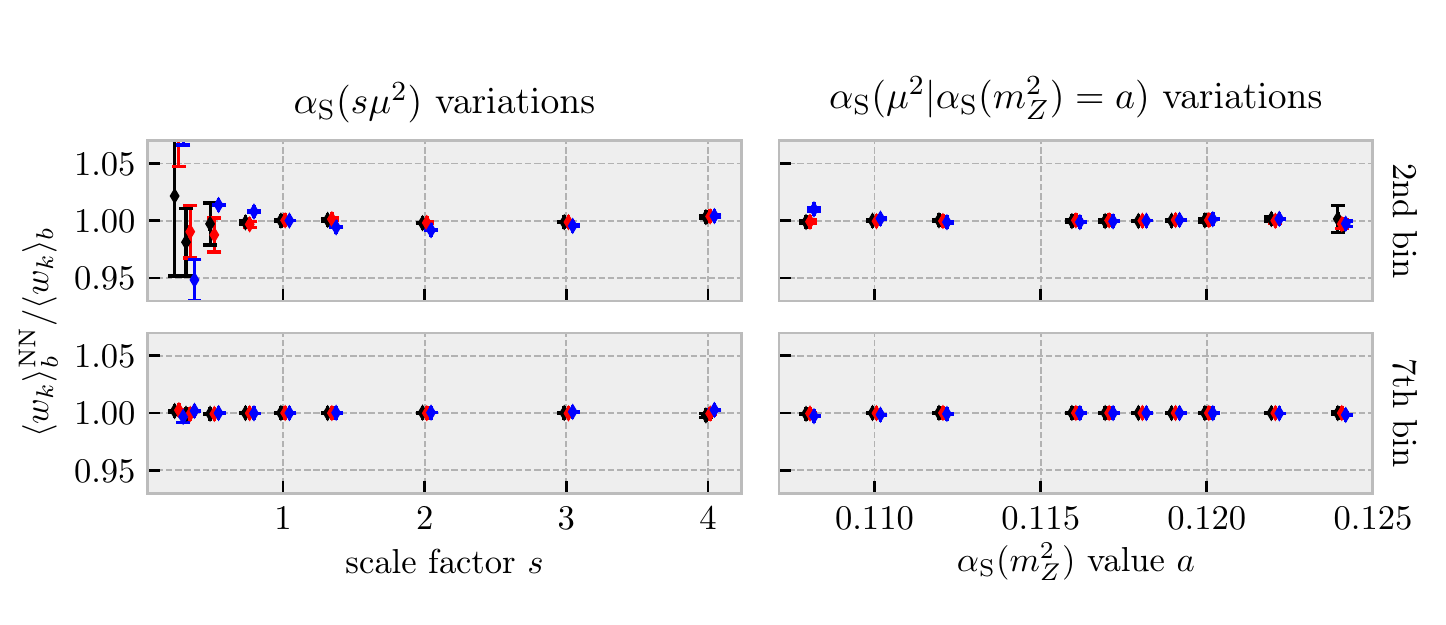}
		\caption{\label{fig:varied_number_of_neurons_study:ptlead}
			leading emission scale $t^\text{lead}$
			}
	\end{subfigure}
	\begin{subfigure}[t]{\textwidth}
		\includegraphics{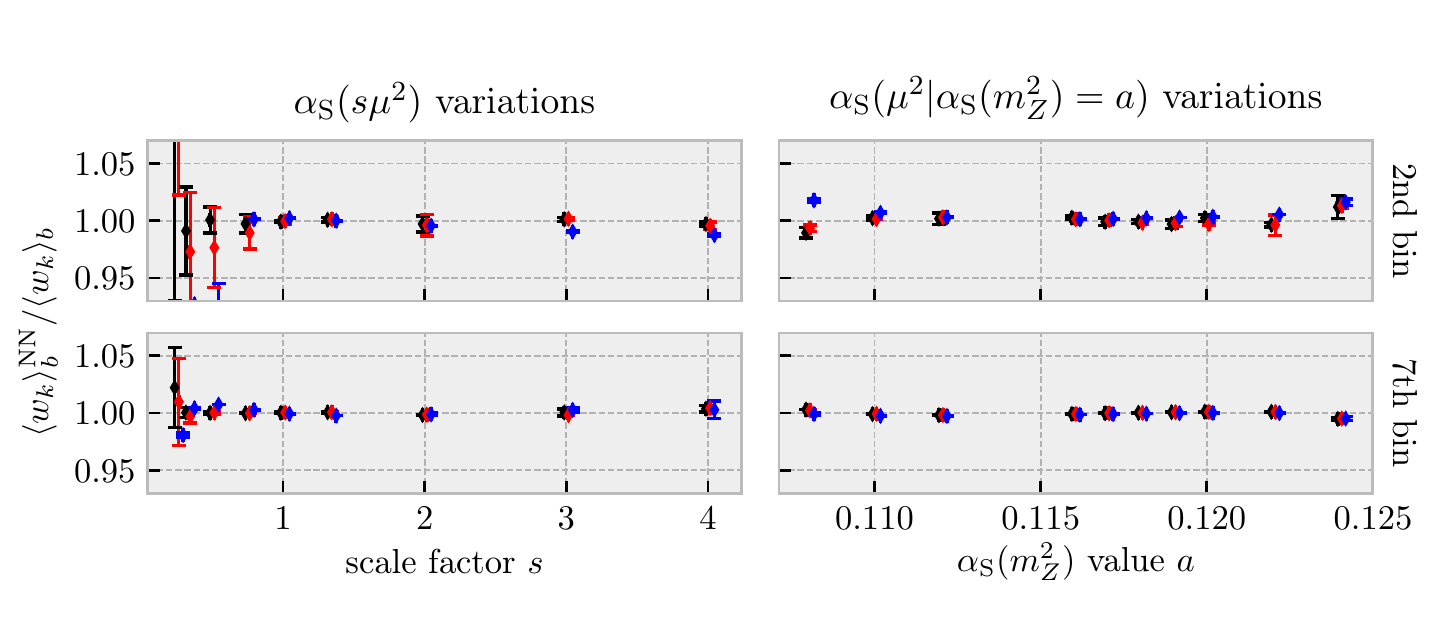}
		\caption{\label{fig:varied_number_of_neurons_study:nemissions}
			number of emissions $N_\text{em}$
			}
	\end{subfigure}
	\caption{
		Ratios between neural-net-predicted and true reweighting factors for a
		low- and a high-$t^\text{lead}$ ($N_\text{em}$) bin are shown in the
		upper (lower) two rows, using $N_\text{in}=60$ (black), 40 (red) and 5 (blue)
		input neurons. The hidden layer of \textsc{ReLU} is in each case set to
		have $N_\text{in}/4$ neurons (rounding down to the next integer).
		}
	\label{fig:varied_number_of_neurons_study}
\end{figure}

Finally, Fig.~\ref{fig:varied_number_of_neurons_study} shows the behaviour of
the neural-network prediction when reducing the number of neurons.
For this study, we return to our original training data set $\mathcal{T}$ that
does not include simultaneous variations. We compare
our previous choice of $N_\text{in}=60$ with using $N_\text{in}=40$ and
$N_\text{in}=5$. The hidden layer always has $N_\text{in}/4$ neurons (which is
rounded down to 1 for the $N_\text{in}=5$ case). Again, we only show two
representative bins for both observables for the sake of brevity. Although
$N_\text{in}=40$ only shows a minor degradation with respect to
$N_\text{in}=60$ it features a substantially increased uncertainty for the
low-$N_\text{em}$ bin and low values of the scale factor $s$. For
$N_\text{in}=5$ we find significant deviations from unity, although even in
this case only the low-$N_\text{em}$ bin features a deviation that is larger
than \SI{5}{\percent} (other than the extremal variations). These findings
suggest that $N_\text{in}=60$ can probably be reduced while preserving enough
precision and accuracy. However, we intend to use the validated architecture in
an example more similar to potential applications in the following section.
For that, we use a full parton-shower implementation, for which we foresee a
stronger variation between the $q_\text{acc}(\mu^2)$ values.
Hence we keep $N_\text{in}=60$ as our baseline architecture.

%% file: text/example.tex
\section{A real-world example: varying Sherpa shower predictions for Thrust}
\label{sec:example}

We now study if the toy shower results from the previous sections can be
transferred to a setup with a complete parton shower implementation
and a real observable, namely the event shape observable Thrust for the
process $e^+ e^- \to \text{2 or more jets}$, simulated at a centre-of-mass
energy of \SI{91.2}{\GeV}.

We generate Monte-Carlo events for this process using the \textsc{Sherpa} event
generator~\cite{Gleisberg:2008ta} and
its Catani-Seymour Shower implementation (\textsc{css})~\cite{Schumann:2007mg}.
Non-perturbative
effects (such as fragmentation and multiple interactions) and electroweak
corrections are disabled. The $e^+ e^- \to q\bar{q}$ matrix element is
evaluated at leading-order in the couplings. The perturbative order of the
running strong coupling is set to include up to two loops. The shower starting
scale is set to the $Z$ mass, i.e.\ $\mu_\text{Q}=m_Z^2=(\SI{91.28}{\GeV})^2$.
The events are analysed using the \textsc{Rivet} analysis
framework~\cite{Buckley:2010ar}.

The set of variations used for the training is listed in
Tab.~\ref{tab:thrust_variations}. Instead of leaving out single training
values as we did for the validation, we use the full training data set
$\mathcal{T}$ and compare the predictions later with several variations that
are not part of $\mathcal{T}$. The
$N_\text{in}=60$ function values for $q_\text{acc}$ used
as the input data are written out from within the \textsc{css} coupling
implementation.
The output data is given by
generating the Thrust distribution for each variation and then calculating the
ratio to the central value ($s=1.0$, $a=0.118$) for each bin. Some of the bins
for lower Thrust values have a sizeable Monte-Carlo error. To take this into
account, we train 20 neural network replicas per bin, and for each training we
generate a
new $\langle w_k \rangle_b$ replica over $k$. For each replica we vary the
$\langle w_k \rangle_b$ values according to their central value and
uncertainty, assuming a Gaussian distribution.

\begin{table}
	\centering
	\caption{List of variations used for training in the example application.}
	\label{tab:thrust_variations}
	\begin{tabular}{@{}llllll@{}}
		\toprule
		\multirow{2}{*}{scale $s$}
			& 0.50 & 0.60 & 0.66 & 0.75 & 0.85 \\
			& 1.15 & 1.33 & 1.50 & 1.70 & 2.00
		\\ \addlinespace
		\multirow{2}{*}{$a=\alphas(m_Z^2)$}
			& 0.108 & 0.110 & 0.112 & 0.114 & 0.116 \\
			& 0.117 & 0.119 & 0.120 & 0.122 & 0.124
		\\ \bottomrule
	\end{tabular}
\end{table}

\begin{figure}
	\centering
	\includegraphics{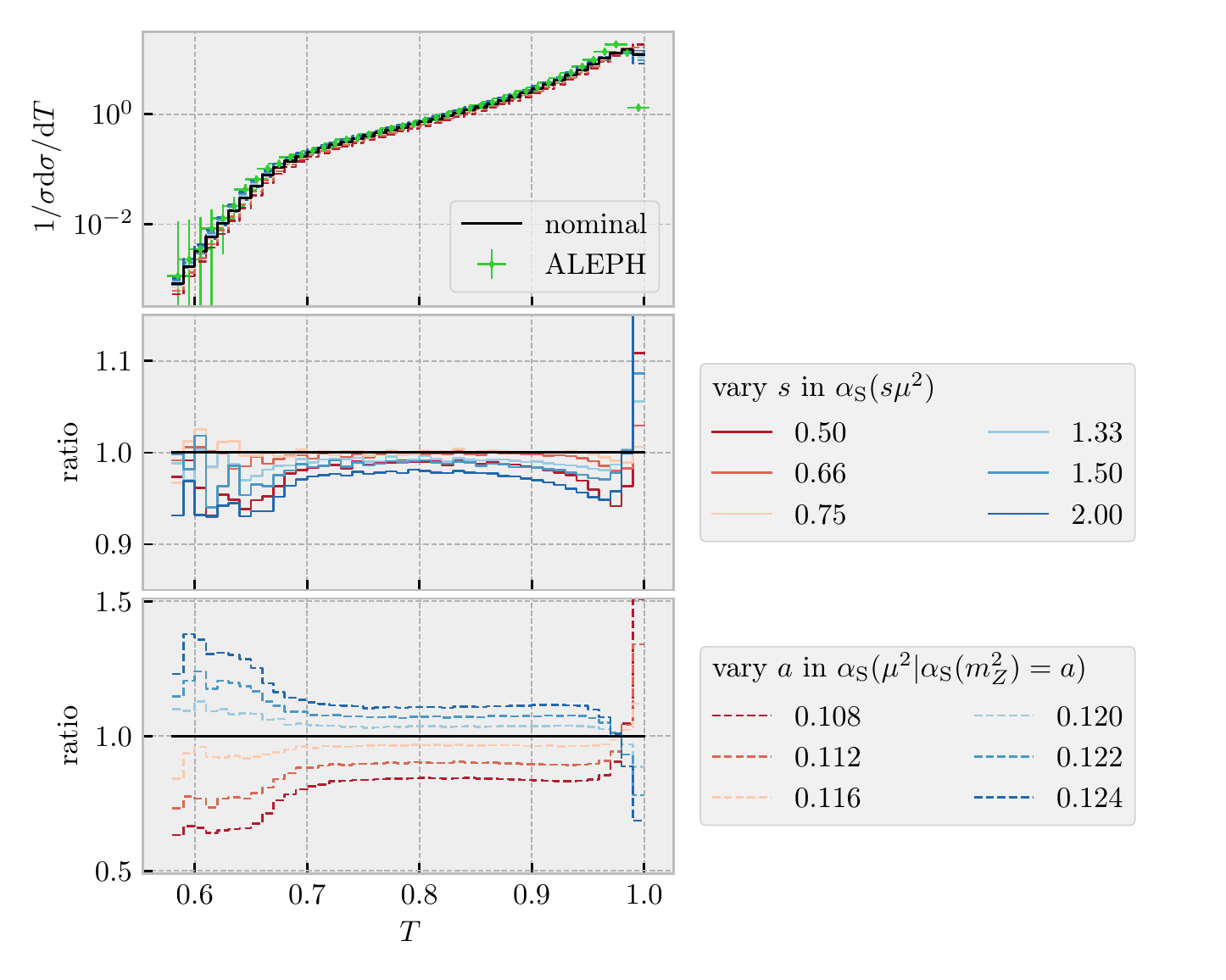}
	\caption{
		The effect of the \textsc{sherpa} Catani-Seymour shower variations on
		the distribution of the Thrust event shape observable. The nominal
		prediction ($a=0.118$, $s=1.0$) is shown in black, whereas the
		variations are colour-coded as listed in the legend on the right. The
		lower panels show the scale and $\alphas(m_Z)$ variations,
		respectively. Some intermediate variations that are used in the
		NN training and comparison are left out to prevent that the
		plots become too busy. In the upper panel, we also show data points
		from the \textsc{aleph}
		collaboration~\cite{Heister:2003aj}.
	}
	\label{fig:thrust_distribution}
\end{figure}

\begin{figure}
	\centering
	\includegraphics{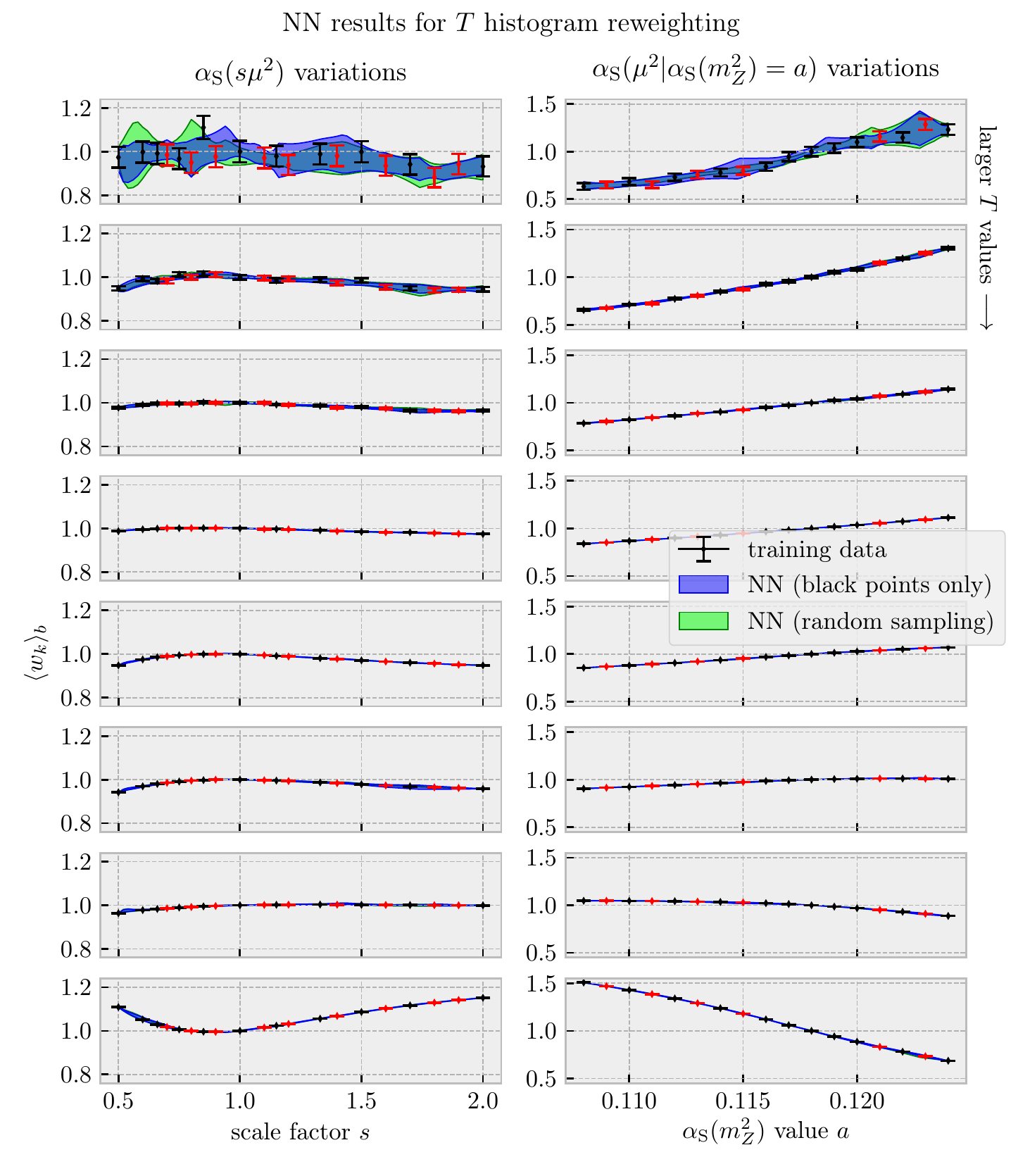}
	\caption{
		A comparison between NN-predicted and true reweighting factors
		for a sample
		of bins of the Thrust $T$ histogram, see
		Fig.~\ref{fig:thrust_distribution}. Each row corresponds to one bin,
		with $T$ being in the intervals (0.58, 0.59), (0.63, 0.64), (0.88,
		0.89), (0.96, 0.97), (0.97, 0.98), (0.98, 0.99), (0.99, 1.00) (from top
		to bottom). The data
		points correspond to the true reweighting factors and their Monte-Carlo
		errors. The black ones are used to train a first set of neural networks
		(blue band). A random sample omitting 7 points for each variation type
		is used to train a second set of neural networks (green band). The most
		extreme variations are always kept in the training set.
		The uncertainties corresponds to the Monte-Carlo error of the training
		data as described in the text.
	}
	\label{fig:thrust_results}
\end{figure}

In Fig.~\ref{fig:thrust_distribution}, we show the LO+parton-shower prediction
for Thrust and a selection of variations for both the scale factor $s$ and for
$a=\alphas(m_Z^2)$, and compare it with data by the \textsc{aleph}
collaboration~\cite{Heister:2003aj}. All reweighting factors for a
representative selection of bins are shown in Fig.~\ref{fig:thrust_results},
along with the prediction for the entire variation ranges by the NN.
This prediction reproduces the reweighting factors that were used to
train the networks (black) and also the factors that are shown as control
points (red). The uncertainties of the prediction follow the Monte-Carlo errors
of the training reweighting factors. We also train a second set of NN,
where we omit for each training pass a random selection of 7~variations in the
training (but keeping the most extremal variations). The resulting band (green)
also reproduces the data.

Note that each NN corresponds to one row in Fig.~\ref{fig:thrust_results}
(i.e.\ to one observable bin), and therefore predicts the
different functional forms for \emph{both} the $a$ and the $s$ variations. The
facts that these functions are non-linear and that their forms depend on the
variation type and observable bin suggest that an ordinary fit with a fixed
parametrisation is not suitable for the task.

%% file: text/conclusions.tex
\section{Conclusions}\label{sec:conclusions}

Parton-shower calculations are currently not included in PDF fits, because of
the CPU time needed to re-evaluate the parton shower event-by-event for new
input parameters, and in particular as the PDFs are changing in the fitting
process. 

In this paper we suggested to use neural networks to encode the
dependences of the cross section in a given observable bin on the parton-shower
input parameters. We showed that the ansatz is working when applied to
variations of the strong coupling (and its input scale) used in the shower
splittings, both for a simplified shower model and for the full shower
implementation in \textsc{sherpa}. The observables we tested are the leading
emission scale, the number of emissions and the Thrust event shape. 

This successful proof of principle makes us confident that it is worth exploring
the method further, to study more observables and variation types but in
particular to generalise it to take into account the PDF dependence of
initial-state shower splittings. This will surely require a more advanced
neural-net architecture or at least considerably more neurons and training data
points, because the dependence of the splitting kernels on the PDFs is more
complicated, and because varying PDF sets can not be done in just one dimension.
It is nonetheless worthwhile trying to implement a fast reweighting procedure
for these processes, in order to extend the range of data that can be used in
PDF fits.

\textbf{Acknowledgements.} We thank Juan Rojo for interesting comments on the manuscript. EB \& LDD are supported by an STFC Consolidated Grant, ST/P0000630/1. LDD is also supported by a Royal Society Wolfson Research Merit Award, WM140078.

%% file: text/sizes.tex
\makeatletter
\newcommand\thefontsize[1]{{#1 The current font size is: \f@size pt\par}}
\makeatother

\section{Sizes}

The text width is: \the\textwidth

\begin{figure}
	\begin{subfigure}[b]{\textwidth}
		\caption{\thefontsize{Subcaption Test}}
	\end{subfigure}
	\caption{\thefontsize{Caption Test}}
\end{figure}